\documentclass{pas}
\usepackage{multirow}
\usepackage{amsmath}
\usepackage{lineno}
\begin{document}
\lefttitle{Publications of the Astronomical Society of Australia}
\righttitle{Owusu et al.}
\jnlPage{1}{13}
\jnlDoiYr{2024}
\doival{10.1017/pasa.xxxx.xx}
\newcommand{\Evans}[1]{\textbf{\color{magenta}Evn: #1}}
\articletitt{Research Paper}

\title{At the same age, metallicity, and alpha-enhancement, sodium is a more effective tracer of the young and old sequences of the Milky Way disc}

\author{\sn{Evans K.} \gn{~Owusu}$^{1,2,3}$, \sn{Sven} \gn{~Buder}$^{2,4}$, \sn{~Ashley J.} \gn{Ruiter}$^{1,2,3,5}$, \sn{Ivo R.} \gn{~Seitenzahl}$^{3}$ and \sn{Nicolas} \gn{~Rodriguez-Segovia}$^{1}$}  

\affil{$^1$School of Science, University of New South Wales Canberra, Australia Defence Force Academy, ACT 2600, Australia \\
$^2$ARC Centre of Excellence for All-Sky Astrophysics in 3 Dimensions (ASTRO-3D)\\
$^{3}$Heidelberger Institut für Theoretische Studien, Schloss-Wolfsbrunnenweg 35, 69118 Heidelberg, Germany\\
$^4$Research School of Astronomy and Astrophysics, Australian National University, ACT 2611, Australia\\
$^{5}$OzGrav: The ARC Centre of Excellence for Gravitational Wave Discovery, Hawthorn, VIC 3122, Australia}

\corresp{E.~K. Owusu, Email: e.owusu@unsw.edu.au}

\citeauth{Owusu et al. 2024, PASA, submitted.}

\history{(Received 01 05 2024; revised 14 06 2024; accepted xx xx xxxx)}

\begin{abstract}
Trends in elemental enrichment with stellar age can give us a powerful avenue to identify thus far unexplained origin sites of the elements. We investigate stellar abundance trends using the GALAH DR3 high-resolution spectroscopic dataset of 6234
 solar-type stars. Our study explores the elemental abundance [X/Fe] of sodium (Na) with stellar age. We find a pronounced enrichment in [Na/Fe] at super solar metallicity (i.e., [Fe/H] $>~0$) in the old sequence of Milky Way disc stars, a trend demanding a deeper understanding of the underlying source(s) responsible for the nucleosynthesis. This progressive [Na/Fe] enrichment at the young end of the old sequence has essential implications for Galactic archaeology. In this work, we propose a novel selection technique for separating the Milky Way's thick and thin disc stellar populations (i.e., old and young sequences) based on the observed [Na/Fe] rise of $\sim$0.1 dex for stars around $5 - 8\,\mathrm{Gyr}$ old. We also compare our selection method to the conventional [Mg/Fe] vs [Fe/H] selection approach, and we find that our new Na-based selection method better disentangles the overlap between young- and old-sequence disc stars at these intermediate ages. This is especially true at super solar [Fe/H], where the [Mg/Fe] vs [Fe/H] or [{$\alpha$}/Fe] vs [Fe/H] separation approaches exhibit significant overlap. This new selection method should help us better understand the history of the formation of the Milky Way disc.
\end{abstract}

\maketitle

\section{Introduction}
\label{sec:Intro} 
Understanding the origin of the elements is a long-standing, unresolved fundamental problem in astrophysics. Numerous works over the last several decades have focused on exploring the production sites \citep[][to name a few]{Burbidge1957,Timmes1995, Jose2011, Kobayashi2020}. Extensive modern spectroscopic surveys like the Galactic Archaeology with HERMES survey \citep [GALAH,][]{DeSilva2015, Buder2018, Buder2021}, provide the best chance to learn more about the synthesis and enrichment of elements in the Milky Way and its satellite galaxies through the measurement of chemical abundances for about \(30\) elements in hundreds of thousands of stars. 
These surveys give us the capability to observe a large number of stars that cover a wide range of Galactic environments, i.e., the Milky Way's stellar halo, bulge and discs, and in some cases, even stars that were born outside the Galaxy \citep [e.g.][]{Buder2022}. Their uniform data acquisition also enables more efficient and reliable data reduction and analysis, thus reducing uncertainties of relative abundances between stars. Such stellar abundance data sets are ideal for evaluating theoretical stellar nucleosynthesis models.

Most of the analysis throughout this paper is focused on the GALAH data for the $^{23}$Na sodium isotope, which in the low-metallicity regime ($-3.0 \leq$~[Fe/H]~$\leq-1$) can be used as a tracer of a star's birthplace, either Galactic or extragalactic \citep{Nissen2010, Smiljanic2016, Das2020, Buder2022,Lombardo2022}. However, the diagnostic potential of Na has not been tested extensively in the metal-rich regime (i.e., above solar metallicity), although the data show promising possibilities. It was noted over 40 years ago that the abundance of Na relative to Fe tends to decrease toward low metallicities \citep{Peterson1981}. A later study by \citet{Adibekyan2012} reported a clear upturn of [Na/Fe]~$> ~0$ (look at the context in the paper). Since then, other studies by \citet{Bensby2014}, \citet{Bensby2017}, \citet{Nissen2020} and \citet{Griffith2022}, have confirmed the Na abundance increment for stars with super-solar metallicity. This is interesting because this metal-rich regime is populated by stars of the young, dynamically cold and thus thin stellar disc. In contrast, the old, dynamically hotter and thick stellar disc is known to be metal-poor. These two populations were first identified in the Milky Way's disc by \citet{Yoshii1982} and \citet{Gilmore1983}, and since then, considerable effort has been put into studying the overlap of these populations \citep[e.g.][] {Bensby2014, Buder2019, Nissen2020}. Until now, no simple, simultaneously observationally- and computationally 
an efficient way to cleanly separate the stars of these two-disc populations has been identified.

\citet{Nissen2020} extended earlier high-precision investigations of elemental abundances in solar twin stars \citep[e.g.][]{DaSilva2012, Nissen2015, Spina2018, Bedell2018} to a wider metallicity range to examine how [Fe/H] influences trends in elemental ratios with stellar age. To determine the ages of 72 nearby solar-type stars with metallicities in the range of \(-0.3 \leq\text{[Fe/H]}\leq 0.3\), \citet{Nissen2020} analysed spectra observed with High Accuracy Radial velocity Planet Searcher (HARPS) with signal-to-noise ratios \(S/N \geq 600\) at \(\lambda \sim 6000~\text{\AA}\). The \AA rhus STellar Evolution Code (ASTEC; \citealt{ChristensenDalsgaard2008}) stellar models were used to determine stellar ages from measured effective temperatures and luminosities. Their study shows that two different populations can be identified namely, an older sequence of stars with a high rise in [Fe/H] to $+0.3$ dex at $7$ Gyr and a younger sequence with an increase in [Fe/H] from $-0.3$ dex to $+0.2$ dex during the previous $6$~Gyr (see their figures~3~and~4). 
Additionally, as a function of stellar age, the trends of some abundance ratios, including $\text{[O/Fe]}$, $\text{[Na/Fe]}$, $\text{[Ca/Fe]}$, and $\text{[Ni/Fe]}$, are divided into two corresponding sequences. On the other hand, the $\text{[Y/Mg]}$-age relation exhibits no offset between the two age sequences and has no appreciable dependence on $\text{[Fe/H]}$. More striking, the age-metallicity distribution being split into two sequences indicates two instances of gas accretion onto the Galactic disc with a quenching of star formation in between. Hereafter, we will follow the notation of \citet{Nissen2020} and refer to the old high-alpha, thick disc as the ``old sequence" and the young low-alpha, thin disc as the ``young sequence".

Similarly, a study exploring the solar neighbourhood model by \citet{Kobayashi2011} found an overproduction of [Na/Fe]. However, a subsequent study \citep{Kobayashi2020} used updated reaction rates and considered non-local thermodynamic equilibrium (NLTE), factors that cause variation towards higher metallicities and affect the production of Na, in turn decreasing [Na/Fe] due to their impact on Type Ia supernovae, which are prominent producers of Fe but do not produce copious amounts of Na 
A recent study by \cite{Cinquegrana2022} predicts an upturn for Na yields at higher metallicities (see their Fig. $4$, bottom right-hand panel). 
The higher-metallicity stellar models of \citet{Cinquegrana2022} are less efficient at achieving and enduring hot bottom burning (HBB) due to the cooler temperatures at the base of the convective envelope. 
When conditions for HBB are achieved, there is some competition between the production of $^{23}$Na via the Ne–Na chain and the destruction of Na through two main channels: $^{23}$Na(p, $\alpha$)$^{20}$Ne and $^{23}$Na(p, $\gamma$)$^{24}$Mg. 
Only the more massive stars (M~$\geq$~7~M$_\odot$) reach temperatures high enough to activate the Ne–Na chain.
The $^{23}$Na surface abundance is correspondingly also found to increase during the second dredge-up (SDU). 

In this paper, we show that the observed upturn of [Na/Fe] at metallicities \(\text{[Fe/H]}~> 0\) appears to be a unique characteristic of the older stellar disc which enables a separation method yielding a less-contaminated sampling of the thin and thick discs. If such a selection is reasonable, one could more efficiently study the stellar populations of the discs when they otherwise overlap in many characteristics like age, metallicity, alpha-enhancement and dynamics. To this end, we use the third data release of the GALAH Survey \citep{Buder2021}, which provides us with stellar chemical abundances and isochrone interpolated ages through the overlap with the \textit{Gaia} satellite \citep{GaiaCollaboration2021}.

In particular, in this study we report on the following:
\begin{itemize}
    \item The [Na/Fe] overproduction in GALAH DR3 at \(\text{[Fe/H]} > 0\) and highlight plausible reasons for it. Additionally, we carry out a brief discussion of several other elements in the context of nucleosynthesis and Galactic enrichment. 

    \item To what degree do our proposed new selection criteria improve upon canonically-used old and young sequence selection methods.
\end{itemize}

Section \ref{sec:galah} describes our data, selection criteria, and methodology for obtaining abundances in GALAH DR3. 
In Section \ref{sec:analysis} we discuss in more detail our standard cut (i.e., see Fig. \ref{fig:cut}) and Section \ref{sec:discussion} presents a brief analysis of the other odd-Z elements, light elements, alpha elements, and elements near the iron-peak. We also show how our new selection method separates the young and old sequence stars as a function of Galactocentric pericentre distance. Finally, in Section \ref{sec:conclusion}, we present a brief conclusion. 

\begin{figure*}
	\includegraphics[width=\textwidth]{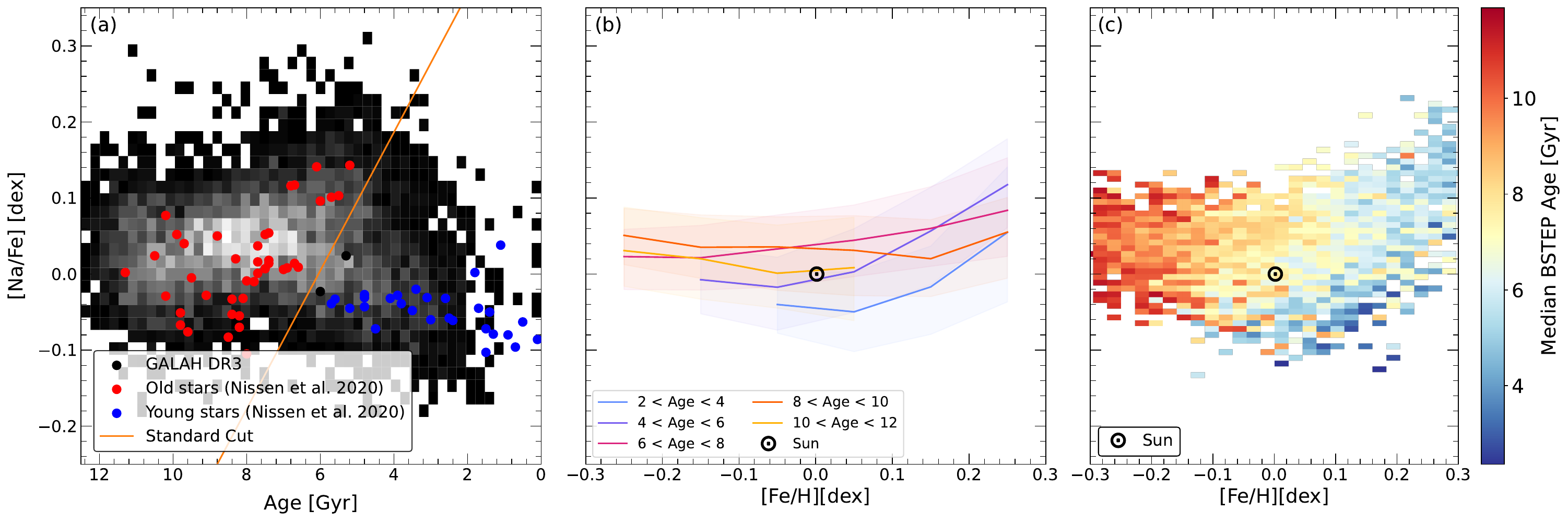}
    \caption{Different representations of the age-iron-sodium abundance relations for solar-type stars of \citet{Nissen2020} and GALAH DR3 (see Eqs.~\ref{eq:solar_type} and \ref{eq:Z_range}). 
    Panel (a) shows age vs. [Na/Fe] measurements by \citet{Nissen2020} with old (red) and young (blue) sequences. The density distribution of GALAH DR3 solar-type stars is shown in greyscale in the background.  Panel (b) shows the same plane but with the distribution of [Na/Fe] in 2 Gyr age bins for 0.1 dex [Fe/H] bins. The lines represent the median, whereas the spread show the 16th to 84th percentiles. Panel (c) shows the 2-dimensional distribution of [Na/Fe] vs [Fe/H] of the GALAH DR3 solar-type stars coloured by median age, with the Sun indicated with $\odot$.}
    \label{fig:age_na_fe_solar_type}
\end{figure*}

\section{Data} 
\label{sec:galah}
\subsection{The GALAH Survey}
The third data release (DR3) of the GALactic Archaeology with High Efficiency and Resolution Multi-Element Spectrograph \citep[HERMES; at the Anglo-Australian Telescope,][]{Sheinis2015}, hereafter GALAH, is a high-resolution (R $\sim$28000) spectroscopic sample of stars in the Milky Way Galaxy containing atmospheric parameters and abundance information for nearly 600\,000 stars as of the current data release \citep{Buder2021}. 

GALAH is an extensive stellar spectroscopic survey like SEGUE \citep{Yanny2009}, RAVE \citep{Kordopatis2013}, and APOGEE \citep{Anders2014}. These surveys serve as a window to study the formation and evolutionary history of the Galaxy and astrophysical processes using traces of gas that have been transported and incorporated into the formation of generations of stars throughout the Galactic history. Primarily, GALAH targets relatively bright stars in the magnitude range of \(9< V < 14\), which are typically closer than $2\,\mathrm{kpc}$. In GALAH DR3, up to 30 elemental abundances are provided for each star, along with their uncertainties and quality flags. Additional physical properties such as surface gravity, effective temperature and age are also reported. The ages were determined using the Bayesian stellar parameter estimation code (BSTEP; \citealt{Sharma2018}). 
 
\subsection{Abundances and how they are determined in GALAH DR3}
GALAH DR3 provides analyses of stellar spectra from the HERMES spectrograph, which covers four wavelength bands in the optical and near-infrared \citep[4\,713 -- 4\,903, 5\,648 -- 5\,873, 6\,478 -- 6\,737, and 7\,585 -- 7\,887~\text{\AA},][]{Sheinis2015}. The abundances were derived using the 1D MARCS model atmospheres \citep{Gustafsson2008}, and a modified version of the Spectroscopy Made Easy (SME) spectrum synthesis code (\citealt{Valenti1996, Piskunov2017}). Of these elements, 11 were computed using TE (\citealt{Amarsi2020}) and the remaining 19 elements were determined using local thermodynamic equilibrium (LTE). Na is among the $11$ elements for which NLTE effects were included in the analysis of GALAH spectra. The abundance analysis is preceded by a step that derives the global stellar parameters from the spectra, namely effective temperature ($T_\text{eff}$), surface gravity ($\log g$), iron abundance [Fe/H], micro-turbulence velocity ($v_{\text{mic}}$), broadening velocity ($v_\text{broad}$) and radial velocity ($v_\text{rad}$). The elemental abundances of element `A' to element `B' are expressed as logarithmic ratios of number densities $N$, that is,
 \begin{equation}
  \left[\text{A}/\text{B}\right] \equiv {\rm log}_{10}(N_\text{A}/N_\text{B}){_\star} - {\rm log}_{10}(N_\text{A}/N_\text{B}){_\odot}
\label{eq:quadratic}
 \end{equation}
 where abundances are normalised to the solar values and are defined in the usual way as [X/Fe] $=$ [X/H] $-$ [Fe/H]. 

The ages of stars in GALAH DR3 were computed using a Bayesian Scheme known as the Bayesian Stellar Parameters estimator \citep[BSTEP][]{Sharma2018} with an elaborate step-by-step distribution given in section $3$ of \cite{Sharma2018}.

\subsection{Sample Selection} \label{sec:sample_selection}
To assess the nature of the abundance enrichment and its correlation with stellar age, we apply an overall quality cut and then divide thea into a smaller sample of solar-type stars, which are defined as cool main-sequence dwarfs located below the red edge of the classical instability strip \citep[including spectral types from late F, G and up to K dwarfs,][]{Soderblom1998,Garcia2019} for the GALAH data, this can be applied through the following masks:

\begin{align} 
    \texttt{flag\_sp} == 0,
    \texttt{flag\_fe\_h} == 0, \texttt{flag\_Na\_fe} == 0, \label{eq:flags} \\
    \texttt{e\_age\_bstep}/\texttt{age\_bstep} < 0.2 \text{ or }\texttt{e\_age\_bstep} < 2, \label{eq:age_cuts} \\
    \texttt{snr\_c2\_iraf} > 50, \label{eq:snr_ccd2}\\
    5600 < T_\text{eff} < 5950, \log g > 4.15     \label{eq:solar_type}, \\
    -0.3 < \mathrm{[Fe/H]} < 0.3.
    \label{eq:Z_range} 
 \end{align}

We employ basic quality cuts to select only unflagged measurements (Eq.~\ref{eq:flags}). Because our study relies on reasonably determined ages, we further limit our sample to stars with either less than $20\%$ age uncertainty or age uncertainties below $2\,\mathrm{Gyr}$ (Eq.~\ref{eq:age_cuts}). To ensure a reasonable quality of measurements, we set a lower spectrum quality cut of signal-to-noise in the green CCD of 50 per pixel (Eq.~\ref{eq:snr_ccd2}).
We further limit our sample to solar-type stars via Eqs.~\ref{eq:solar_type} and \ref{eq:Z_range} following the definition of \citet{Nissen2020}.

Whenever we use elemental abundances [X/Fe], we further apply a quality cut of
\begin{equation} 
    \texttt{flag\_X\_fe} == 0
\end{equation}
where X replaces the respective element, e.g. Na, for acceptable sodium measurements. 

\section{Analysis}\label{sec:analysis}

The motivation for this project is the \citet{Nissen2020} age-metallicity separation (see their figs.~3~and~4) of old and young solar-type stars, which have been historically equated as thick and thin disc sequences, respectively. The upturn in [Na/Fe] with age for the old solar-type stars in their measurements especially caught our eye. We reproduce this plot in Fig.~\ref{fig:age_na_fe_solar_type}a, identify an upturn of Na in the [Na/Fe] vs [Fe/H] (Fig.~\ref{fig:age_na_fe_solar_type}c) and show the[Fe/H] versus [Na/Fe] relationship with age bin in step size of 2 Gyr (see Fig.~\ref{fig:age_na_fe_solar_type}b) for $\mathrm{[Fe/H]} > 0$ in the GALAH data. In Fig.~\ref{fig:age_na_fe_solar_type}c), the standard deviation of the ages for stars around solar metallicity is $\pm 2$ Gyr. 

Our analysis first focuses on the dissection of the upturn of [Na/Fe] at iron abundances above $\mathrm{[Fe/H]} > 0$ in Sec.~\ref{sec:analysis_na_upturn}. We then show that increasing [Na/Fe] vs age can improve the selection of co-natal young and old sequence stars in Sec.~\ref{sec:analysis_selection}. In Sec.~\ref{sec:analysis_age_xfe_relations}, we then assess the trends of young and old sequences in the other age-abundance planes available with GALAH DR3 data before confirming the robustness of our selection function. 

\begin{figure*}
	\includegraphics[width=\textwidth]{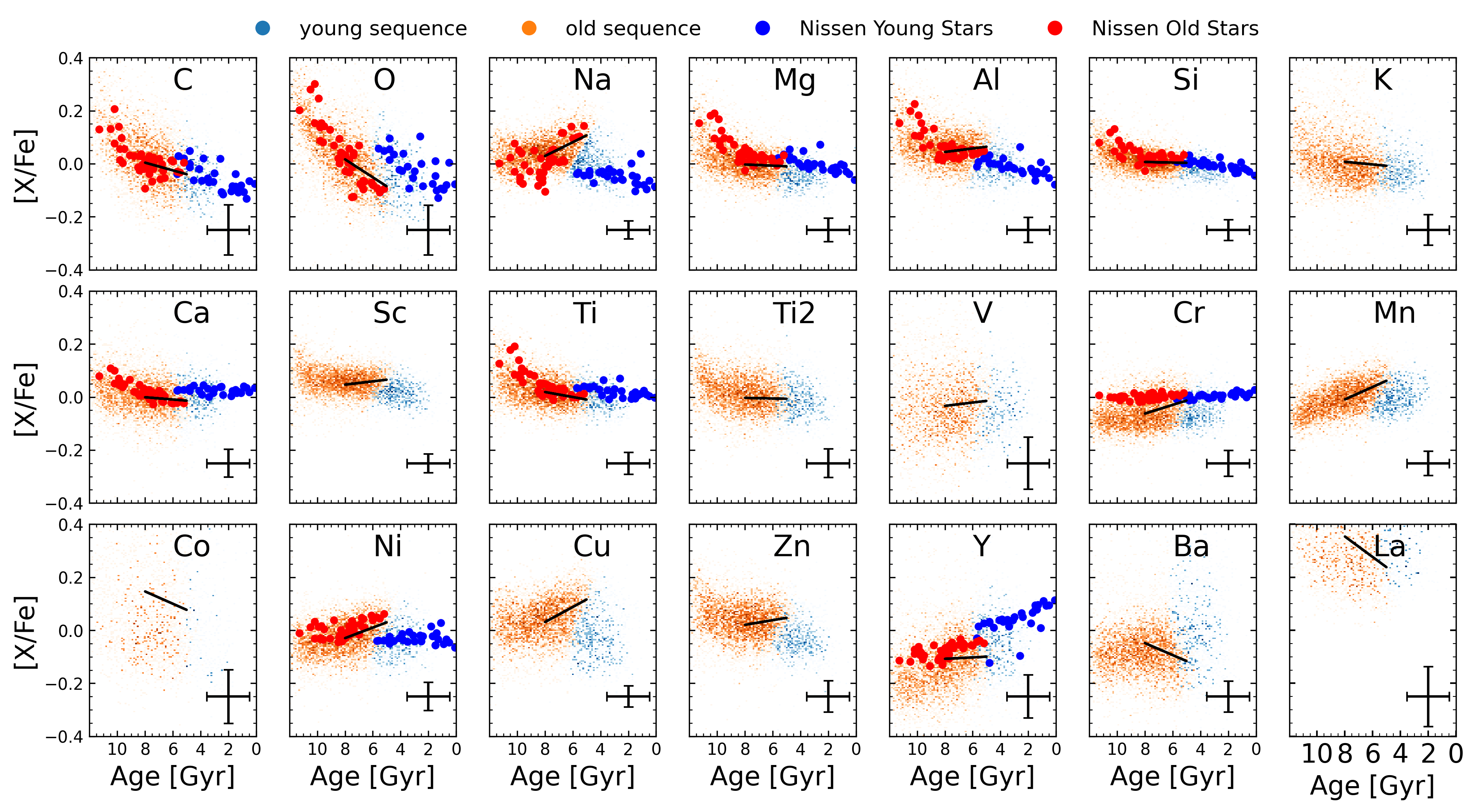}
    \caption{Age-abundance distributions of young and old solar-type stars for the elements measured by GALAH DR3 (small dots in the background) and \citet[][larger dots in foreground]{Nissen2020}. Reddish colours represent stars of the old sequence and blueish colours represent stars of the young sequence, respectively, as selected via Eq. \ref{eq:line_age_na_fe}.
    The elements are indicated in each panel. The black lines indicate a linear fit to the old sequence of GALAH DR3 between $5$ and $8\,\mathrm{Gyr}$. Typical uncertainties in this age range are shown as error bars in the bottom right.}
    \label{fig:age_abundance}
\end{figure*}

\begin{figure}
	\includegraphics[width=\columnwidth]{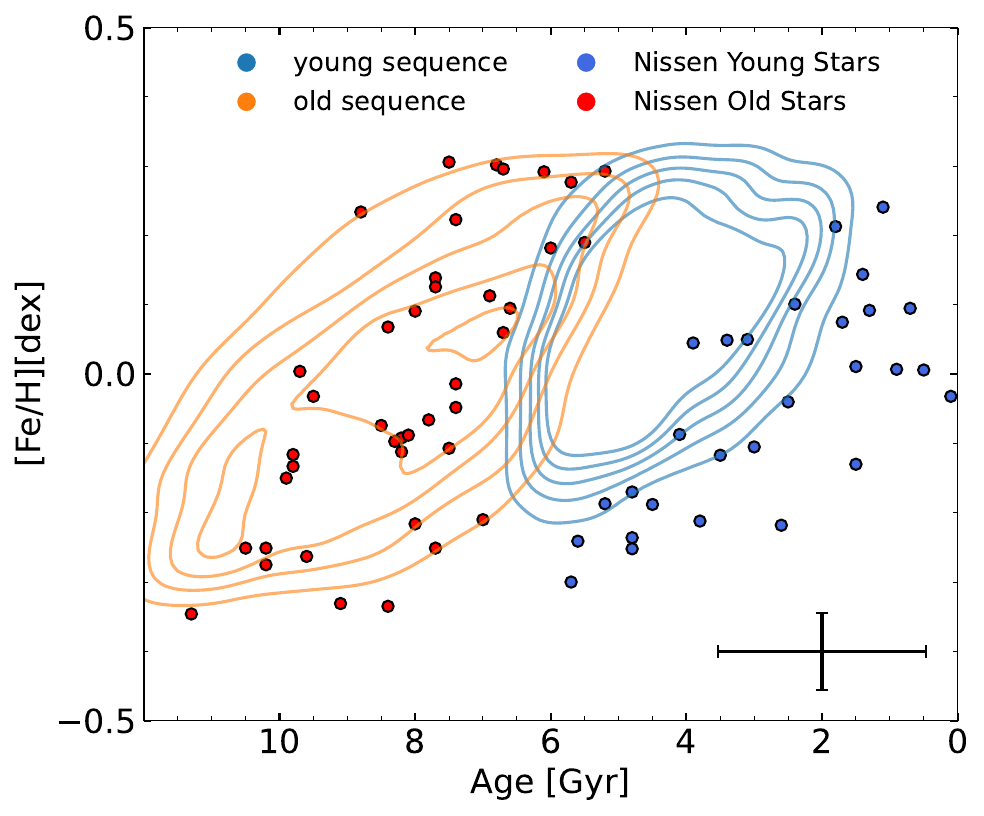}
    \caption{The distribution of stars in the [Fe/H]-age plane for both older and younger sequences of solar-type stars in GALAH DR3 is shown using contours. The contour levels are $0.1, 0.2, 0.3, 0.4,$ and $0.5,$ with each level representing the probability of selecting a star belonging to a specific classification. The contour levels are increasing inwards. The larger dots represent stars from the old and young sequence as per \citet{Nissen2020} and they are colour-coded according to the classification defined in Figure \ref{fig:age_abundance}.}
    \label{fig:contour}
\end{figure}

\subsection{The upturn of [Na/Fe] at [Fe/H]$~>~0$} \label{sec:analysis_na_upturn}

In Fig. \ref{fig:age_na_fe_solar_type}b) we show the [Na/Fe] versus [Fe/H]  relation with bin median ages in a step size of $2$ Gyr. 
The important takeaway is that within all age bins explored in Fig. \ref{fig:age_na_fe_solar_type}b, stars are visibly more enriched in [Na/Fe] at high metallicity for ages between 4 -- 6 Gyr. 
The resulting differences in [Na/Fe] enhancement for different ages motivated us to test its potential to separate stars of the young and old sequence of the Milky Way disc.

\subsection{Selecting young and old stars in the [Na/Fe]-Age plane} \label{sec:analysis_selection}
The Milky Way's stellar disc is a complex structure, and its formation and evolution are still widely debated \citep[e.g.][]{Katz2021}. Several surveys have provided large statistical samples of stars with detailed chemical abundances, such as RAVE \citep{Kunder2017}, Gaia-ESO \citep{Randich2013}, APOGEE \citep{Alam2015}, LAMOST \citep{Zhao2012}, and GALAH \citep{Buder2018}, combined with very precise Gaia astrometry \citep{Lindegren2018}, making it possible to examine the formation of the Galactic disc with an unprecedented level of detail. The sample includes mainly thin and thick disc stars of all ages selected based on various criteria, such as the relative abundance of alpha-elements and stellar ages within the old sequence of the Milky Way disc. Due to the overlap of the thin and thick disc stars, several studies \citep[see for instance][]{Lee2011, DelgadoMena2021, Bensby2014, RecioBlanco2014, Hayden2017} have devised preliminary techniques to separate the contamination of the two-disc sequences. However, kinematics and chemo-dynamics have proven to be rather inaccurate in distinguishing both sequences, especially at intermediate ages \citep{Hayden2017} in the metallicity range of $-0.3$ to $+0.5$ dex, where both sequences overlap \citep[e.g.][]{Bensby2014}.

 The Milky Way's stellar disc is a complex structure, and its formation and evolution are still widely debated \citep[e.g.][]{Katz2021}. Several surveys have provided large statistical samples of stars with detailed chemical abundances, such as RAVE \citep{Kunder2017}, Gaia-ESO \citep{Randich2013}, APOGEE \citep{Alam2015}, LAMOST \citep{Zhao2012} and GALAH \citep{Buder2018}, combined with very precise Gaia astrometry \citep{Lindegren2018}, making it possible to examine the formation of the Galactic disc with an unprecedented level of detail. The sample includes mainly thin and thick disc stars of all ages selected based on various criteria, such as the relative abundance of alpha-elements and stellar ages within the old sequence of the Milky Way disc. Due to the overlap of the thin and thick disc stars, several studies \citep[see for instance][]{Lee2011, DelgadoMena2021, Bensby2014, RecioBlanco2014, Hayden2017} have devised preliminary techniques to separate the contamination of the two-disc sequences. However, kinematics and chemo-dynamics have proven to be rather inaccurate in distinguishing both sequences, especially at intermediate ages \citep{Hayden2017} in the metallicity range of $-0.3$ to $+0.5$ dex, where both sequences overlap \citep[e.g.][]{Bensby2014}.

We continue our analysis by reproducing the selection of young and old sequences in the age-abundance plane similar to \citet{Nissen2020}. However, contrary to \citet{Nissen2020}, who use a selection in the age vs. [Fe/H] space, we select stars via age and [Na/Fe]. Thick and thin solar-type stars from the study of \cite{Nissen2020} 
are superimposed over the GALAH DR3 solar-type stars data (see Fig. \ref{fig:age_na_fe_solar_type}).

Fig \ref{fig:age_na_fe_solar_type}a, in particular, shows an initial selection (hereafter standard cut) based on a separation `by eye' of the \citet{Nissen2020} stars as an orange line. We classify stars as young or old whether they are below or above the line as:
\begin{equation}
\left.
\begin{aligned}
\text{old stars} & : \text{[Na/Fe]}_{\text{old}} \,\,\,\,\,\,> 0.55 - 0.5/5.5 ~\times\tau \\
\text{young stars} & : \text{[Na/Fe]}_{\text{young}} < 0.55
 - 0.5/5.5~\times\tau
\end{aligned}
\right\},
\label{eq:line_age_na_fe}
\end{equation}

where $\tau$ represents the stellar age. We discuss the sensitivity of our results to variations of our standard cut in Sec. \ref{sec:analysis_age_xfe_relations} in which we test the robustness of our selection criteria for the GALAH data sample. For the smaller data set of\citet{Nissen2020}, this cut is robust mainly due to the difference of [Na/Fe] rise of $\sim~0.30$ dex around $5\,\mathrm{Gyr}$.

Our selection now allows us to assess the age-abundance distribution for all other elements in Fig.~\ref{fig:age_abundance}, which shows the solar-type thick (red colour) and thin (royal blue colour) disc stars from \citet{Nissen2020} overlaid on GALAH DR3 solar-type thick (orange) and thin disc (sky blue) stars respectively. Further to this, we show the distribution of our stellar sample population with contours in the [Fe/H]--age plane in Fig. \ref{fig:contour}. The GALAH DR3 solar-type population in our work follows a similar trend in separation in the [Fe/H] vs. age space as depicted in \cite{Nissen2020}. Since our dataset comprises nearly 2 orders of magnitude more stars than the \citet{Nissen2020} sample, it includes stars at about $6.3$ Gyr, the age that roughly corresponds to the transition between young and old sequence and is sparsely populated in the \cite{Nissen2020} work (see Section \ref{sec:6p5_gyr_break} for further discussion). We note that the ages of younger stars tend to be larger than those of \citet{Nissen2020}, but the trends agree in a relative sense, which thus does not change the implications of this study. Recent works e.g. \citet{Xiang2022} and \citet{Anders2023} have confirmed the existence of two sequences in the [Fe/H]--age diagram. 

While we have established the clear upturn of [Na/Fe] for the old sequence, the trends for other elements are more diverse. While other odd-Z elements like aluminium (Al) and scandium (Sc) also demonstrate a rise in their abundance with decreasing stellar age, their rise is less pronounced than that of sodium (Fig. \ref{fig:scatter)}. When we consider potassium (K), another odd-Z element, it shows a unique trend (i.e., it remains flat), though we note K is less well-constrained in the [X/Fe] axis. Furthermore, some alpha elements (i.e., C, O, and Ca) show overall decreasing trends as we consider younger stellar ages; see also Tab. \ref{tab:age_percentiles}. On the other hand, elements near the iron peak (i.e., Cr, Mn, Ni, Cu, Co and Zn) demonstrate increasing upward trends. More so, the neutron capture element yttrium (Y) remains flat, while barium (Ba) shows a decreasing trend. Section \ref{sec:discussion_nucleosynthesis} discusses possible underlying reasons for these trends.

\begin{figure*}
	\includegraphics[width=\textwidth]{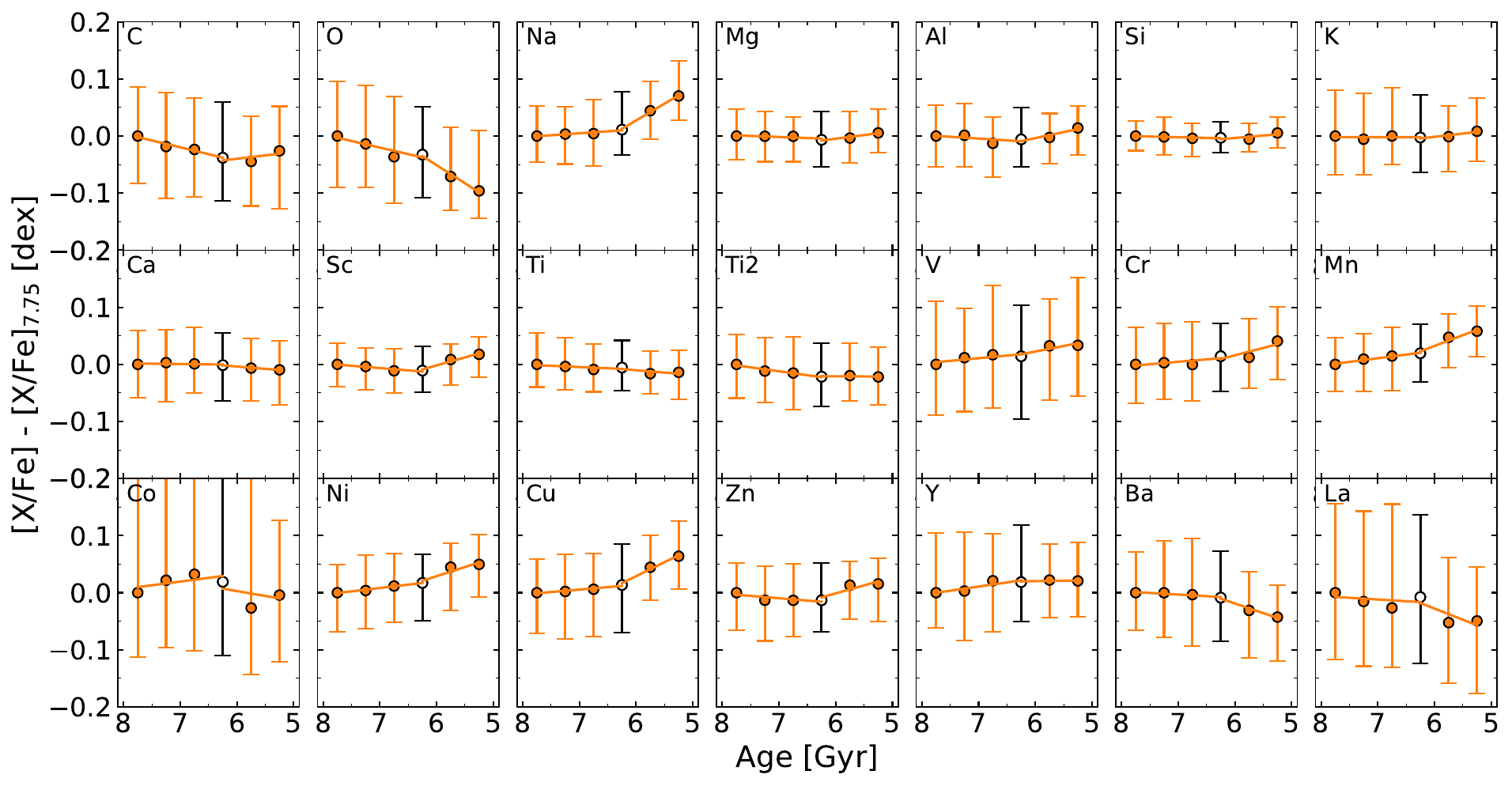}
\caption{Age-median abundance trends of the old-solar type stars with ages 5 -- 8 Gyr as reported in GALAH DR3. Each panel shows its corresponding element on the top-left corner, and scatter dots represent the median abundance in a $0.5$ Gyr width bin. Uncertainties are taken from the abundance distribution in a given age bin, with the lower (upper) uncertainty equalling the 16th (84th) percentile. For visualisation purposes, all median abundances have been shifted so the value at the first bin ($\sim 7.75$ Gyr) has solar abundance. An empty marker is used at $6.25$ Gyr to highlight the position at which we find a break in the linear trend for elements such as C, Na and Sc (see Section \ref{sec:6p5_gyr_break}}). The linear fit for stars above/below this limit are shown as solid orange lines and motivate the results shown in Table \ref{tab:age_percentiles}, where we report the results for linear fits to the individual stars above/below $6.3$ Gyr and also the entire 5 -- 8 Gyr age range.
\label{fig:scatter}
\end{figure*}

\begin{figure}
	\includegraphics[width=\columnwidth]{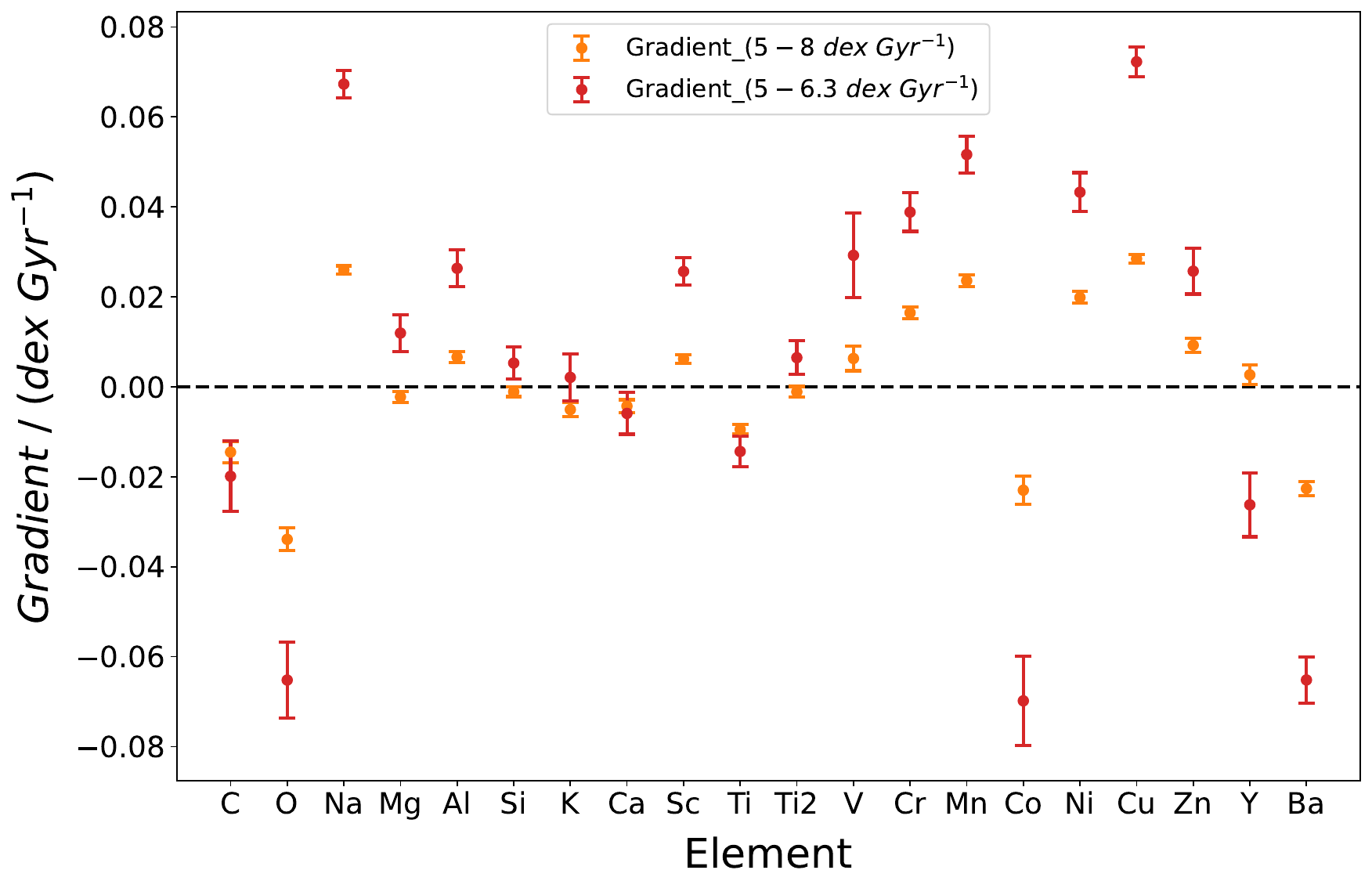}
    \caption{
    Gradient estimates of abundances, $\Delta \mathrm{[X/Fe]}/\Delta \mathrm{Age}$ of different elements X for old solar-type stars as selected with the Standard Cut from Eq.~\ref{eq:line_age_na_fe}. 
    Positive values indicate an increase of element X over iron with time, whereas negative values show a relative decrease in the abundance of element X relative to iron with time. Error bars indicate the fitting uncertainties. We note that most of the enrichment for intermediate mass and iron peak elements occurs between $6.3$ and $5$ Gyr ago, e.g. during the birth of stars representing the younger end of the old sequence as given in Tab. \ref{tab:age_percentiles}}
    \label{fig:combined_slopes_by_element}
\end{figure}

\begin{table*}
    \centering
    \caption{We report the gradient for the entire 5 -- 8 Gyr age range, as well as the trend for stars younger/older than $6.3$ Gyr motivated by Fig. \ref{fig:scatter}. Note that in this case, the linear fits were performed on the individual stars and not on the binned abundances, since the ages within a bin are not uniformly distributed and therefore using a representative age value (as done in Figure \ref{fig:scatter}) would introduce unnecessary uncertainties. A graphical representation is given in Fig. \ref{fig:combined_slopes_by_element}}
    \begin{tabular}{ccccccc}
\hline \hline
Atomic Number & Element & Gradient $5.0-8.0\,\,\mathrm{Gyr}$  & Gradient $5.0-6.3\,\,\mathrm{Gyr}$ & Gradient $6.3-8.0\,\mathrm{Gyr}$ \\
& & $\mathrm{dex\,Gyr^{-1}}$ & $\mathrm{dex\,Gyr^{-1}}$ & $\mathrm{dex\,Gyr^{-1}}$\\
\hline
6             & C           & $-0.0145 \pm 0.0024$               & $-0.0199 \pm 0.0078$             & $-0.0125 \pm 0.0053$               \\
8             & O           & $-0.0339 \pm 0.0025$               & $-0.0652 \pm 0.0084$             & $-0.0210 \pm 0.0056$               \\
11            & Na          & $\,\,\,\,0.0260 \pm 0.0009$        & $\,\,\,\,0.0673 \pm 0.0031$      & $\,\,\,\,0.0066 \pm 0.0021$        \\
12            & Mg          & $-0.0022 \pm 0.0012$               & $\,\,\,\,0.0119 \pm 0.0041$      & $-0.0090 \pm 0.0027$               \\
13            & Al          & $\,\,\,\,0.0066 \pm 0.0012$        & $\,\,\,\,0.0263 \pm 0.0041$      & $-0.0113 \pm 0.0027$               \\
14            & Si          & $-0.0011 \pm 0.0011$               & $\,\,\,\,0.0053 \pm 0.0035$      & $-0.0068 \pm 0.0024$               \\
19            & K           & $-0.0051 \pm 0.0016$               & $\,\,\,\,0.0021 \pm 0.0053$      & $-0.0058 \pm 0.0035$               \\
20            & Ca          & $-0.0043 \pm 0.0014$               & $-0.0059 \pm 0.0046$             & $-0.0017 \pm 0.0031$               \\
21            & Sc          & $\,\,\,\,0.0062 \pm 0.0009$        & $\,\,\,\,0.0256 \pm 0.0031$      & $-0.0052 \pm 0.0021$               \\
22            & Ti          & $-0.0095 \pm 0.0011$               & $-0.0144 \pm 0.0035$             & $-0.0103 \pm 0.0024$               \\
22            & Ti2         & $-0.0011 \pm 0.0012$               & $\,\,\,\,0.0065 \pm 0.0037$      & $-0.0054 \pm 0.0029$               \\
24            & V           & $\,\,\,\,0.0063 \pm 0.0027$        & $\,\,\,\,0.0292 \pm 0.0094$      & $-0.0019 \pm 0.0062$               \\
25            & Cr          & $\,\,\,\,0.0164 \pm 0.0013$        & $\,\,\,\,0.0388 \pm 0.0043$      & $\,\,\,\,0.0092 \pm 0.0029$        \\
27            & Mn          & $\,\,\,\,0.0236 \pm 0.0012$        & $\,\,\,\,0.0516 \pm 0.0041$      & $\,\,\,\,0.0137 \pm 0.0027$        \\
28            & Co          & $-0.0230 \pm 0.0031$               & $-0.0698 \pm 0.0099$             & $\,\,\,\,0.0544 \pm 0.0077$        \\
29            & Ni          & $\,\,\,\,0.0199 \pm 0.0013$        & $\,\,\,\,0.0433 \pm 0.0043$      & $\,\,\,\,0.0064 \pm 0.0029$        \\
30            & Cu          & $\,\,\,\,0.0285 \pm 0.0010$        & $\,\,\,\,0.0723 \pm 0.0034$      & $\,\,\,\,0.0080 \pm 0.0023$        \\
39            & Zn          & $\,\,\,\,0.0092 \pm 0.0015$        & $\,\,\,\,0.0257 \pm 0.0051$      & $-0.0076 \pm 0.0034$               \\
56            & Y           & $\,\,\,\,0.0027 \pm 0.0022$        & $-0.0262 \pm 0.0071$             & $\,\,\,\,0.0160 \pm 0.0048$        \\
57            & Ba          & $-0.0226 \pm 0.0016$               & $-0.0652 \pm 0.0052$             & $-0.0016 \pm 0.0035$               \\ \hline \hline
\end{tabular}
    \label{tab:age_percentiles}
\end{table*}

\subsection{Quantifying age-abundance trends and testing the newly suggested selection via the age-[Na/Fe] plane} \label{sec:analysis_age_xfe_relations}

Here we test how robust our age-abundance trends are regarding the particular selection function (cut guided by eye) chosen in the [Na/Fe]-age plane. The results of the age-abundance trends for each element are displayed in Fig. \ref{fig:scatter}. The outcomes of the linear fits performed on our population of stars are presented in Table \ref{tab:age_percentiles}. 

\begin{figure}
	\includegraphics[width=\columnwidth]{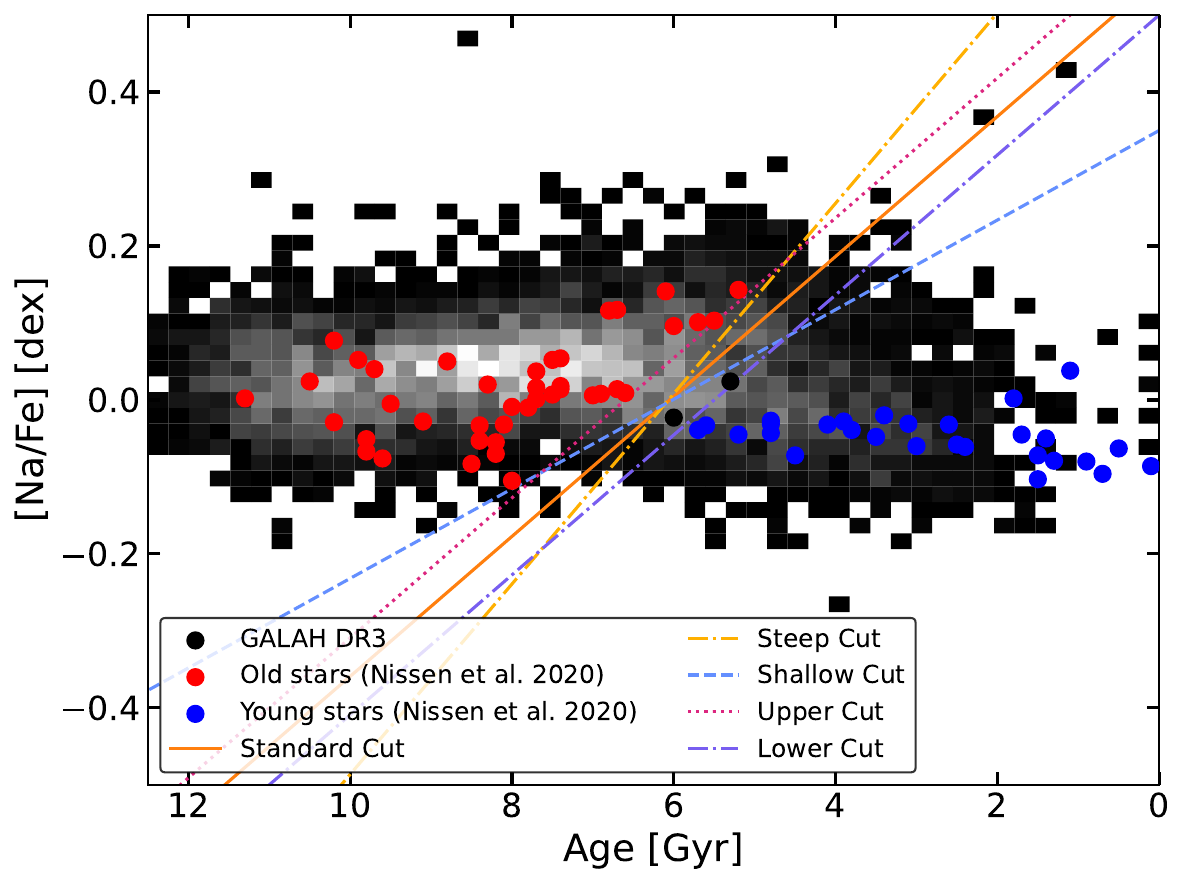}
     \caption{Similar to  Fig.~\ref{fig:age_na_fe_solar_type}a for the entire age range considered (8--5 Gyr), but now including different cuts to separate old from young solar-type stars (see Eqs.~\ref{eq:upper_cut}-\ref{eq:shallow_cut}).}
    \label{fig:cut}
\end{figure}

\begin{figure}
	\includegraphics[width=\columnwidth]{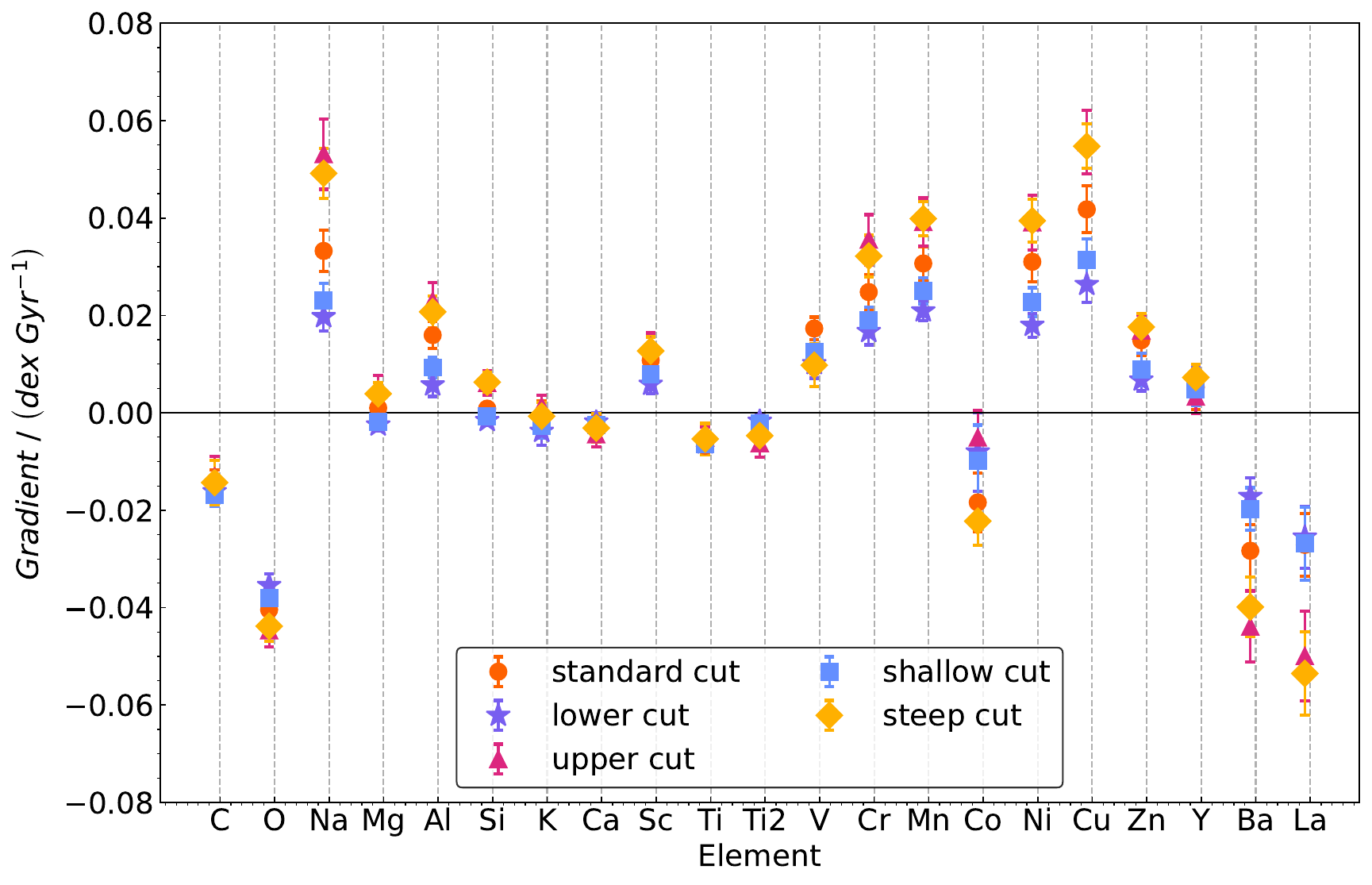}
    \caption{Gradients for the old sequence stars from this work, similar to   Fig.~\ref{fig:combined_slopes_by_element}, but this time, the gradient value is taken over the larger age range that is shown in Fig. \ref{fig:cut} for the various selection cuts that are displayed. Colours correspond to those used in Fig.~\ref{fig:cut}. We note that modifying the slope or moving the y-intercept up or down does not significantly change our findings, e.g. the gradient values remain mostly clustered.}
    \label{fig:gradient}
\end{figure}

To test the robustness of our [Na/Fe] versus age trend against the details of the choice of the standard cut in Eq. \ref{eq:line_age_na_fe}, we consider further cuts (as shown in Fig. \ref{fig:cut}) defined by the equations below:\\

\begin{align}
\text{upper cut} & : \text{[Na/Fe]} = 0.55\,(-0.05) - 0.5/5.5 ~\times\tau, \label{eq:upper_cut} \\
\text{lower cut} & : \text{[Na/Fe]} = 0.55\,(+0.05) - 0.5/5.5~\times\tau, \label{eq:lower_cut} \\
\text{steep cut} & : \text{[Na/Fe]} = 0.75 - 0.68/5.5~\times\tau\text,
\label{eq:steep_cut} \\
\text{shallow cut} & : \text{[Na/Fe]} = 0.35 - 0.32/5.5~\times\tau,
\label{eq:shallow_cut}
\end{align}
where $\mathrm{\tau}$ is the age of stars in Gyr. The effect of these various cuts is shown in Fig. \ref{fig:gradient}. 

It has been noted that cobalt (Co) has large uncertainty values because only a few lines are present in GALAH DR3. However, for other elements, the trends obtained from different separation cuts agree with the scatter, meaning that the slopes characterising abundance relative to iron with decreasing stellar age all cluster around the same gradient value for the elements investigated. This suggests that the slope estimates for different cuts of each component are in close agreement, indicating a robust selection technique for separating the old sequence from the young sequence.
\section{Discussion}\label{sec:discussion}

Now that we have established a correlation between [Na/Fe] enhancement with decreasing stellar age for the Milky Way old sequence solar-type stars, 
we discuss both the implications on the nucleosynthetic channels at play and what we could learn from this about the thin and thick disc. As the main feature of our analysis, we first discuss the age-abundance relation of Na (Sec.~\ref{sec:discussion_na_age}) in detail before including a discussion about other elements in Sec.~\ref{sec:discussion_nucleosynthesis}. This will inform our discussion of the nucleosynthetic implications for the old sequence in Sec.~\ref{sec:discussion_implications_nucleosynthesis}. Having focused on the old sequence, we will then put our results into the context of the young disc in Sec.~\ref{sec:discussion_young_sequence}. Given the overlap of young and old discs around $6.3\,\mathrm{Gyr}$, we further discuss the nature of the change in age-abundance trends for our old sequence selection in Sec.~\ref{sec:6p5_gyr_break}, before discussing how much the selection of the two sequences improves when including [Na/Fe] in Sec.~\ref{sec:discussion_improvement_seleciton}.

\subsection{The age-abundance trend of Na for the old sequence} \label{sec:discussion_na_age}

Sodium is an odd-Z element synthesised in exploding massive stars and dying low-mass stars. In massive stars, Na is synthesised during the hydrostatic carbon burning phase \citep{Cameron1959, Salpeter1952,Woosley1995}. Also, during the neon-sodium (Ne-Na) and magnesium-aluminium (Mg-Al) cycles in dying low-mass stars, Na is produced \citep{Denisenkov1990} and mixed to the surface (e.g. \citealt{ElEid1995, Mowlavi1999, Karakas2010, Cinquegrana2022}).

Owing to the diverse sources from which Na is synthesised, its abundance pattern as a function of metallicity is considered a complex one \citep{Bedell2018}. For instance, the trend of [Na/Fe] first increases with [Fe/H] because of the contributions from massive stars that produce Na and Fe \citep[][see fig 21]{Bensby2017}; however when the nucleosynthetic products of SNe~Ia begin to have a significant impact on a stellar population ${\sim}0.4 - 1$ Gyr after star formation \citep[][see fig. 2]{Ruiter2011}, [Na/Fe] starts to drop with [Fe/H]. As a result, the trend of [Na/Fe] over time (i.e., metallicity and age) has been described as a ``zigzag" one \citep{McWilliam2016}. This trend is important for our consideration because the bulge shares the same abundance trend with the thin and thick discs \citep{Bensby2017}. 
The study of \citet{Nissen2020} reported a rise in [Na/Fe] of $0.15$ dex for the same stellar population at the same age range (5 -- 8 Gyr). Similarly, we find that a linear fit to the old sequence at intermediate ages of $5 - 8$ Gyr gives a gradient value of $0.0260 \pm 0.0009\,\mathrm{dex/Gyr}$
, i.e, $0.077$ dex for the stated age range for [Na/Fe]. 
Also, a steeper rise for [Na/Fe] occurs between the ages of $5 - 6.3$ Gyr with an abundance gradient of $0.0673 \pm 0.0031\,\mathrm{dex/Gyr}$
 (see red data points in Fig \ref{fig:combined_slopes_by_element}) compared with that of $0.0066 \pm 0.0021\,\mathrm{dex/Gyr}$
 (see Tab. \ref{tab:age_percentiles}). 

We indeed see an explicit ``knee'' behaviour \citep{McWilliam2016} for several elements in our study, with a plateau behaviour for stars aged 8-6.3 Gyr followed by a change in slope -- going up or down -- from ages 6.3 to 5 Gyr. One may assume then that iron enrichment from SNe~Ia becomes more critical around the time of the ${\sim} 6.3$ Gyr knee. 
We can see a decline for some elements relative to iron as we observe younger (generally more metal-rich) stars owing to the (generally long) delay times attributed to SNe~Ia \citep{Matteucci1990,Ruiter2009, Maoz2012}. An upward trend in [X/Fe] with decreasing age would be expected for some elements, particularly around the iron peak \citep[i.e. Mn, which is mainly synthesized in SNe~Ia arising from massive white dwarf stars close to the Chandrasekhar mass limit,][]{Seitenzahl2013a}. However, Na also shows an upward trend with decreasing age, though Na is not canonically associated with SNe~Ia but more often with short-lived, massive stars. This means something is driving Na-enrichment other than simply the products of massive stars. The knee effect we see in Na here would be consistent with Na being synthesised by core-collapse supernova progenitors with a metallicity-dependent yield -- the `metallicity effect' -- which we describe in the following three paragraphs.  


Many of the light odd-Z elements only have one stable isotope (e.g., $^{23}\text{Na}$ (11p,12n), $^{27}\text{Al}$ (13p,14n), $^{31}\text{P}$ (15p,16n), $^{45}\text{Sc}$ (21p,24n), etc.). These odd-Z stable isotopes are neutron-rich with nuclei having more neutrons than protons. At low progenitor metallicity, a pre-supernova star will have $Y_e$ very close to 0.5 in the relevant layers where explosive nucleosynthesis occurs since core helium burning produces mostly self-conjugate nuclei like $^{12}\text{C}$,$^{16}\text{O}$, and $^{20}\text{Ne}$. However, with increasing metallicity, more C, N, and O nuclei are already present during hydrogen burning. The CNO cycle turns most of the nuclei present as CNO to $^{14}\text{N}$. 

Importantly, a main reaction during core helium burning then is
$^{14}\text{N} (\alpha,\gamma) ^{18}\text{F} (\beta^+ \nu_e) ^{18}\text{O} (\alpha,\gamma) ^{22}\text{Ne}$. In effect, by number, most CNO nuclei initially present are turned into $^{22}\text{Ne}$, which, owing to the beta-decay of $^{18}\text{F}$ is neutron-rich with two extra neutrons. Since the relative number of protons to neutrons is largely conserved for many environments where explosive nucleosynthesis occurs, the extra neutrons available allow the more efficient nucleosynthesis of the neutron-rich odd-Z isotopes. 

In contrast, alpha elements (i.e., the even-Z elements such as Mg, S, or Ca) have more stable isotopes that extend from self-conjugate stable nuclei on the alpha-chain [e.g., $^{24}\text{Mg}$ (12,12), $^{32}\text{S}$ (16,16), $^{40}\text{Ca}$ (20,20)] to several neutron-rich stable isotopes (e.g. $^{25}\text{Mg}$, $^{26}\text{Mg}$, \(\ldots, ^{48}\text{Ca}\)). This means we can make stable even-Z element isotopes at high and low (no) neutron excess. Still, we can synthesise odd-Z elements more efficiently in environments with extra neutrons (which means higher metallicity). Higher progenitor metallicity here also means that the neutron-rich isotopes like $^{25}\text{Mg}$ of these even-Z elements are produced in greater abundance; however, this comes at the expense of the also stable $^{24}\text{Mg}$ and so the total production of the element Mg is hence overall less affected. 

Another potential source of Na-enhancement to reasonably consider is Novae. Novae are (recurring) thermonuclear eruptions on accreting white dwarfs, and could also be considered as a plausible source of Na given their range of delay times \citep[see fig 12,][]{Kemp2021}. However, preliminary work indicates that novae are not a leading contributor to sodium enrichment in the Galaxy above solar metallicity (A. Kemp, Private Comm. 2024).

\subsection{Age-abundance trends other than Na for the old sequence} 

\label{sec:discussion_nucleosynthesis}

\subsubsection{Light elements} 
\paragraph*{Carbon (C),} is synthesised in exploding massive and dying low-mass stars. The trend in [C/Fe] is mainly due to the reduced impact of core-collapse SNe due to the short life-cycle of $\lesssim 10^7$ years for high-mass stars, and [C/Fe] remains flat compared to other light elements in the older sequence of the analysed stellar population (5--6.3 Gyr). Also, the carbon gradient does not change exceptionally across the different age ranges where the fits were conducted (i.e., 5--8, 6.3--8, ~\text{and}~ 5--6.3 Gyr). 

\paragraph*{Oxygen (O)} enjoys the distinction as the most abundant nucleus synthesised in stars because it is the natural end product of helium burning. The majority of O is synthesised in massive stars, partly in dying low-mass stars, with some contribution also from exploding white dwarfs (WD) \citep{Timmes1995, Roepke2012}. The contributions of SNe~Ia and exploding WDs to the synthesis of O are relatively small compared to core-collapse supernovae.  
 
Oxygen is one of the elements where the gradient of [O/Fe] vs age sharply changes around 6.3 Gyr (see Fig.~\ref{fig:scatter}). This is reflected in the fact that the gradient for the young end in the age range of 5.0 to 6.3 Gyr) of the old sequence is much steeper than the gradient for the whole time window in the range of 5 to 8 Gyr, see~Fig.~\ref{fig:combined_slopes_by_element}).   
 This trend indicates a decline in oxygen over iron production within the old sequence in this time frame, perhaps due to SNe Ia producing iron peak elements in the younger generation of old disc stars. The decrease is consistent with the prevalent notion \citep[e.g.][]{Timmes1995} that [O/Fe] decreases with decreasing age as SNe~Ia progressively becomes more dominant and an essential source of iron. Furthermore, we observe what appears to be a knee-like feature in the data consistent with previous findings by (\citealt{McWilliam1997, Nissen2020}). This feature is typically understood to represent the process of gas accumulating onto the thin disc, accompanied by a temporary slowdown or pause in star formation activity during this transition period \citep{Nissen2020}.

\subsubsection{Alpha elements}
Alpha-elements are thought to originate and be expelled back into the interstellar medium predominantly from core-collapse supernovae. Several alpha-elements also have considerable contributions from low-mass stars. As a result, they are valuable indicators of the chemical development of the Galaxy \citep{Bensby2017}.
From Fig. \ref{fig:combined_slopes_by_element}, we see that the alpha elements (Mg, Si, Ca, and Ti) are consistent with no significant changes to their abundances in the 5--8 Gyr range. There is a slight change for Mg and Si around 6.3 Gyr, where the gradient slightly turns upwards (see Fig.~\ref{fig:combined_slopes_by_element}), and Ca remains essentially unchanged.

\paragraph*{Magnesium (Mg)} is synthesised primarily during explosive carbon-burning in exploding massive stars \citep{Arnett1995, Woosley1995} as well as type Ia supernovae \citep{Thielemann2002,seitenzahl2013b}. On average, core-collapse supernovae eject about ten times as much Mg as Type Ia supernovae, and core-collapse supernovae also occur about five times more frequently \citet{Li2011}. From Fig. \ref{fig:scatter}, Mg has an initial flat plateau with a transition to an increasing trend at around 6.3 Gyr. We note that this is different from the conventional knee effect seen in the solar neighbourhood that describes a decrease of alpha/Fe with age as low mass stars (mostly via Fe-group producing thermonuclear supernovae) begin to modify the abundance trends \citep[e.g.][]{Bensby2017}. Notably, if anything, we see a slight increase in [Mg/Fe] in the old sequence around 6.3 Gyr towards younger stars. It is widely accepted that SNe~Ia produces sub-solar [Mg/Fe] \citep[e.g.][]{Timmes1995, Seitenzahl2013a, Lach2020}, making this upward trend difficult to explain with an increasing contribution from thermonuclear supernovae. 

\paragraph*{Silicon (Si)} exists primarily due to oxygen-burning \citep{Woosley1986}. Simply put, two $^{16}$O nuclei collide, causing the reaction
$^{16}$O$ \,\,+\, ^{16}$O $ \longrightarrow \, ^{28}$Si$ \,\,+ \,^{4}$He.
 In addition, some Si is synthesised in exploding WDs and dying low-mass stars \citep{Matteucci1986, Timmes1995}. We see that the abundance of Si is constant, i.e., its slope (against time) is flat like Ti and other alpha process elements (e.g., Mg, Ca) with a slight increase in abundance at 6.3 Gyr (see Tab. \ref{tab:age_percentiles};  
Fig. \ref{fig:combined_slopes_by_element}). Also, the trend in the older sequence of the younger stellar population (from 6.3 Gyr) for [Si/Fe] slightly increases with age.
The discontinuity occurring at ${\sim}6.3$ Gyr could be indicative of a different star formation rate, or gas accretion/enrichment phase, between the old and young sequence discs (\citealt{McWilliam1997, Nissen2020}).

\paragraph*{Calcium (Ca)} is primarily synthesised owing to oxygen-and silicon-burning processes (both explosive and non-explosive) in massive stars \citep{Timmes1995}. The age vs. abundance trend for Ca across all ages exhibits mostly unchanged across all ages. 

\paragraph*{Titanium (Ti)} is primarily synthesised through the alpha-rich freezeout in core-collapse supernovae and explosive helium burning processes \citep[see][]{Tur2010, Kromer2010, Crocker2017}. In GALAH DR3, Ti2 refers to abundances derived from ionised Ti lines (all other alpha-elements are measured from their neutral lines); Table \ref{tab:age_percentiles} lists the gradients for the age-abundance trend both. The age-abundance trend for Ti remains flat in the old sequence. 

\subsubsection{Odd-Z elements}
Odd-Z elements (Na, Al, K and Sc) can provide crucial information about nucleosynthetic processes and chemical evolution in the early Galaxy. When non-LTE effects are considered, abundance trends for odd-Z elements display different trends when compared to the case when NLTE effects are not taken into account \citep{Cayrel2004, Bandyopadhyay2020}. The age-abundance trend of the younger populations (ages between 5-6.3 Gyr) of the old sequence for the odd-Z elements increases more than other elements, with Na having the most pronounced rise (and being the main focus of this work). 


\paragraph*{Aluminium (Al)} is mainly synthesised during carbon-neon (C-Ne) burning in massive stars and the magnesium-aluminium (Mg-Al) cycle in massive AGB stars (i.e., 9 - 12 M$_{\odot}$; \citealt{Siess2006, Karakas2010}). The gradient of Al for the younger population of the older sequence (6.3--5 Gyr) is approximately four times steeper than the trend for the full range of the old sequence (8--5 Gyr). 

More so, even though Al is an odd-Z element, we found a behavioural trend similar to most $\alpha-$elements: a plateau in the age range of 6.3--8 Gyr before an increase that gives rise to a visible knee effect at ${\sim}6.3 $ Gyr. An Al-enhancement at younger stellar ages has been noted in previous studies for stars in the bulge by \citet{Bensby2014, Bensby2017}. 

\paragraph*{Potassium (K)} is primarily synthesised during the burning of oxygen in massive stellar explosions \citep{Woosley1995,Timmes1995, Cayrel2004}. K exhibits a mostly flat trend compared to other odd-Z elements (Na, Al, and Sc). 

While the different production sites for K compared with Na and Al could contribute to the observed trend difference, again, we caution that uncertainty values are relatively high for potassium, particularly at young stellar ages (see Tab. \ref{tab:age_percentiles}).

\paragraph*{Scandium (Sc)}
is mainly synthesised in core-collapse SNe through oxygen burning and the alpha-rich freeze-out reassembly of alpha particles close to the core. Factors such as stellar rotation \citep{Prantzos2018} and two-dimensional (2D) jet-induced supernovae \cite{Kobayashi2020} can lead to increased production of Sc. From \ref {fig:scatter}, Sc remains relatively flat until around 6.3 Gyr where its abundance starts to rise. Furthermore, from Tab.\ref{tab:age_percentiles}), the gradient shows a clear increase as we go to younger stellar ages.

\subsubsection{Elements near the iron peak}
\label{subsec:ironpeak}

Elements near the iron peak (V, Cr, Mn, Ni, Cu and Zn)\footnote{Though Co is an iron-peak element, it shows a “knee-like” trend. We
note that cobalt measurements have comparatively large errors, so we are currently unable to quantify the validity of this apparent trend} follow the expected predictions from nucleosynthesis. Their abundance increases over time due to the increasingly important role of SNe~Ia, which produce large amounts of these elements (in addition to iron) compared to other sources. From Fig. \ref{fig:scatter}, an ``elbow'' effect (a trend opposite to a knee effect), is apparent, which could indicate the onset of different star formation rates between the old and young sequences. Such an effect is seen around 6.3 Gyr for the following elements: Cr, Mn, Ni, Cu and Zn. 
While Cr, Mn and Ni (and also V) show a somewhat steady increase in abundance toward younger ages, Cu and Zn exhibit the most prominent elbow effect, where the behaviour from near-flat to strongly increasing abundance is quite clear. 

\paragraph*{Vanadium (V)} is produced both in massive stellar explosions \citep{Weaver1978} and thermonuclear SNe \citep{seitenzahl2023a}. Vanadium's most abundant stable isotope is $^{51}$V, with one dominant source plausibly being helium-shell detonations in sub-Chandrasekhar mass white dwarf SNe~Ia \citep[][see also \citet{Panther2021} for prospects on detection of V lines in double-detonation SNe~Ia]{Timmes1995}. 
Taken over the entire age range considered for the old sequence (i.e., 5 -- 8 Gyr), the V abundance steadily increases; however, upon closer inspection, we can see that the gradient over the younger part of the age range ( 5 -- 6.3 Gyr) is quite a bit steeper (see also Tab. \ref{tab:age_percentiles}).

\paragraph*{Chromium (Cr)} is predominantly synthesised during incomplete silicon burning in SNe~Ia from near-Chandrasekhar mass white dwarfs \citep{seitenzahl2017a} and explosive oxygen and silicon burning more generally (see \citet{Cote2020,Bergemann2010}).The more neutron-rich isotopes of Cr (such as $^{54}$Cr) are produced in high-density Chandrasekhar mass SNe~Ia, where electron captures drive the material to be more neutron-rich. This was argued by \citet{Bravo2022} as further support for the idea that the majority of SNe~Ia must come from sub-Chandrasekhar (lower density) white dwarfs. 
Our work shows a rise in the abundance of Cr relative to iron, similar to the behaviour of other Fe-peak elements. More so, more Cr is produced in the age range 6.3--5 Gyr as we see a rise in the gradient by a factor of ${\sim}4$ compared to the older 6.3--8 age range. 

\paragraph*{Manganese (Mn)} is mostly synthesised through Type Ia supernovae and, therefore, it has become a critical element for inferences on the nature of SNe Ia from chemical evolution studies, both in the Milky Way \citep{Seitenzahl2013a, Lach2020, Eitner2020, Eitner2023} as well as in Dwarf Galaxies \citep{north2012, mcWilliam2003, mcWilliam2018, deLosReyes2020}. The creation of Mn in near-Chandrasekhar mass SNe Ia depends crucially on the entropy during freeze-out from nuclear statistical equilibrium \citep[for reviews see][]{seitenzahl2017a,seitenzahl2023a}. For sub-Chandrasekhar mass models, Mn production mainly occurs in incomplete silicon burning, where the outcome depends on the metallicity of the explosion  \citep{Lach2020}. In this study, we observe a consistent enrichment and a noticeable increase in the 5 -- 6.3 Gyr age range (see Fig. \ref{fig:combined_slopes_by_element}). 

\paragraph*{Nickel (Ni)} is mostly synthesised by SN~Ia explosions with roughly ${\sim}1/3$ of the contribution coming from core-collapse supernovae \citep{Woosley1995, Timmes1995} in both explosive silicon burning and through alpha-rich freeze-out. The increase in Ni observed for younger stellar populations has also been observed in \citet{Feltzing2009} and \citet{Nissen2020}.

\paragraph*{Copper (Cu)} is largely thought to be produced predominantly through exploding massive stars. From Fig. \ref{fig:combined_slopes_by_element}, we see a relatively high Cu enrichment in the young population of the old sequence between $6.3$ and $5$ Gyr. The first account of the overproduction of Cu was in the seminal study by \citet{Sneden1991}. Also, according to \citet{Bisterzo2004}, the synthesis of Cu in massive stars is attributed to a weak neutron capture process during hydrostatic He-burning and C-burning phases in massive stars. In addition, \citet{Weaver1978} and \cite{Iwamoto1999} predict a high abundance of copper at solar metallicity, citing the overproduction due to the metallicity effect from the progenitors of core-collapse SNe (see \ref{sec:discussion_na_age}). Canonically, it has been predicted that Cu is mostly synthesised through exploding massive stars \citep{Woosley1995}. As a result,  at higher metallicities, the expectation is that the abundance of Cu would decrease. However, our result shows a significant increase (see Tab. \ref{tab:age_percentiles}). Although Type Ia supernovae are generally not considered to make a substantial contribution to the production of Cu, \citet{Lach2020} found from their theoretical approach that the helium shell detonation model of SNe~Ia produces high Cu relative to Fe. Taken together, these findings consequently challenge the existing paradigm for important nucleosynthesis sites of copper. 

\paragraph*{Zinc (Zn)} is synthesised primarily in massive and dying low-mass stars \citep{Nomoto1997,Mishenina2002,Nomoto2013}.

Similar to other elements around the iron peak, we see a change in the abundance gradient as a function of age around 6.3 Gyr, (see \ref{subsec:ironpeak}). However, we note from Fig. \ref{fig:scatter} that Zn initially exhibits a somewhat flat, but slightly decreasing trend, after which we see a rise in its abundance (the elbow effect) around 6.3 Gyr, which is similar to the trend we observe in Al and the $\alpha$ element Mg (and possibly Si).  

 \subsubsection{Others}
Both yttrium (Y) and barium (Ba) are neutron-capture elements produced through nucleosynthesis in dying low-mass stars. We discuss briefly our interpretations below. 
\paragraph*{Yttrium (Y)} is a neutron capture element mainly synthesised in AGB stars \citep{ Bensby2017,Kobayashi2020}. At higher temperatures, Y suffers from NLTE effects, i.e., the lines are weakened due to increasing [Fe/H] or increasing effective temperature $T_\mathrm{eff}$ of the stellar model atmosphere \citep{DelgadoMena2021, Bensby2014, Bensby2017, Storm2023}. We note that the abundance of Y is well below solar over the old sequence (see Fig. \ref{fig:age_abundance}) and remains rather flat (though overall rising) over the entire old sequence age range. 
 
\paragraph*{Barium (Ba)}is one of the elements synthesised through dying low-mass stars. From Tab. \ref{tab:age_percentiles}, we find that the difference in the gradient in the 8--6.3 Gyr and 6.3--5 Gyr age range varies by a factor of forty. We note the older age range of the old sequence exhibits a rather flat trend (perhaps a slight decrease) up to about 6.3 Gyr (Fig. \ref{fig:scatter}). This particular trend is unique among all elements studied in this work, though alpha elements like Ca and possibly Ti arguably show a similar trend. 

\subsubsection{Key findings for age-abundance trends}

We found the following overall trends among the element groups:
\begin{itemize}
    \item Strong, positive enhancement trends for most odd-Z elements (Na, Al, Sc) and most elements near the iron peak (V, Cr, Mn, Ni, Cu, Zn),
    \item A relatively flat abundance trend for alpha-process elements (Mg, Si, Ca, and also Ti),
    \item A decreasing trend for O, as well as C (but less so) on the light side and Ba on the heavy side of the periodic table,
    \item For Co and La, our results were inconclusive. We attempted to explain possible reasons for the behaviours exhibited by K and Y, though we note that  the uncertainty for these elements is relatively high. 
\end{itemize}

\subsection{Implications for nucleosynthesis} \label{sec:discussion_implications_nucleosynthesis}

The increasing trend of [(Na, Al, Sc)/Fe] differs from what we expect based on their canonically assumed primary nucleosynthetic origins. One potential factor leading to an increasing abundance of Na with decreasing stellar age could be the effect of increasing metallicity on Galactic enrichment from core-collapse supernova explosions \citep[][see \ref{sec:discussion_na_age}]{Woosley1994, Kobayashi2006, Nomoto2013,Limongi2018}. Another could be asymptotic giant branch (AGB) stars; AGB stars have been proposed as a possible reason for the rising abundance trend at super-solar metallicity (\citealt{Shi2004, Kobayashi2011, Cinquegrana2022}). 

For most Fe-group elements, the observed rising trend is consistent with predictions for the existing nucleosynthetic paradigm due to the increasingly dominant role of exploding white dwarfs with time (i.e. as we look at younger stars in the Galaxy, we increasingly see the footprint of SN~Ia nucleosynthesis). Furthermore, $\alpha-$elements (i.e., Mg, Si, Ca, and additionally Ti) generally remain flat for the old sequence between ages of $\sim 8 - 6.3$ Gyr age range, after which there is a gradual decrease (so-called knee; first reported by \citet{McWilliam1997}) in their elemental abundances due to the diminishing contribution of core-collapse SNe and increasing birthrate of SNe~Ia. The trend in magnesium (Mg) follows the same age-abundance trend seen in all iron-peak elements in the younger old disc sequence (with a positive gradient), contrary to what we may expect (i.e. that Mg should follow a similar behaviour as other $\alpha$ elements relative to iron and flatten out or become more negative with decreasing age). Also, we find that, light elements (C and O), as well as the neutron capture element Ba show a decreasing trend towards younger stellar ages.


Of the elements that showed enrichment in their abundances, of copper is most prominent, followed by sodium (see Tab. \ref{tab:age_percentiles}). On average, for these elements, more enrichment occurred at around 6.3 Gyr and this manifests as an increase by a factor of $\sim 10$ in their respective gradients. 
Although copper (atomic number $=29$) and zinc (atomic number $= 30$) lie close on the periodic table, they have different age-abundance gradient values. Cu and Zn abundance trends are somewhat flat in the $8$ and $6.3$ Gyr age group. However, Zn quickly rises and remains flat in the $6$ and $5$ Gyr age range. On the other hand, Cu has a gradual but sustained increase in its abundance from ${\sim}6$ Gyr.

\subsection{Putting the old sequence trends into the perspective of the young sequence}\label{sec:discussion_young_sequence}
While our study focuses on the old sequence of the Milky Way, a similar survey for solar-twin stars belonging to the thin disc was carried out by \citet{Spina2016} to trace the chemical evolution by examining the [X/Fe] versus age for $24$ elements from carbon (C) to europium (Eu). \cite{Spina2016} used the high-resolution and high signal-to-noise UVES optical spectrograph of the VLT to obtain spectra of nine solar twins to make precise estimates of stellar ages and chemical abundances. The study revealed that each class of elements showed a distinct evolution with time, depending on the different characteristics, rates, and timescales of the nucleosynthesis sites from which they are produced. Findings from the study showed that $\alpha$-elements (Ca, Ti, Mg, Si, etc.) are characterised by a [X/Fe] decrease with age, while the iron-peak elements (V, Cr, Mn, Ni, Co, etc.)  show an early [X/Fe] increase that peaks at around $6$ Gyr followed by a decrease towards the youngest stars. In addition to previous work by \citet{Spina2016} on Solar twin stars, \cite{Buder2021} compared the Solar twin stars in GALAH DR3 to that of \citet{Spina2016} and found a significant agreement for these elements: O, Na, Si, Ca, Sc, Ti, Mn, Zn, Y, and Ba. However, elements such as Mg and Cr had a slight offset due to few lines being used in the calibration red {\citep[see Fig.~15 of][]{Buder2021}}. The presence of few lines in our data for Cr is the reason for the offset when compared with other elements that agree with that of \citet{Nissen2020} data.

\subsection{The significance of an enrichment change at 6.3 Gyr} \label{sec:6p5_gyr_break}
We appreciate that, at first glance, we have made a seemingly arbitrary cut when fitting the age-abundance trends above and below 6.3 Gyr. However, this decision is also corroborated by previous reports of a knee-age effect by \citet{Edvardsson1993} and \citet{McWilliam1997}. Since then, other studies have reported this knee-age effect, mostly seen in $\alpha-$element plots \citep{Bensby2014, Spina2016}. The occurrence of the knee indicates the onset of the dominant role of (ideally) one nucleosynthesis source relative to another. Various authors have provided different explanations for the presence of the knee phenomenon. One view, proposed by \citet{McWilliam1997}, suggests that the knee effect indicates varying star formation rates between distinct stellar populations such as the bulge, halo, and Galactic discs. Another view, expressed by \cite{Nissen2020}, suggests that the knee can be interpreted as evidence of gas accretion onto the Galactic disc, with a quenching of star formation in between. 

Our work shows the occurrence of the knee-age effect (or elbow-age effect) in most elements considered around 6.3 Gyr. For odd-Z and Fe-peak elements (see Section \ref{sec:discussion_nucleosynthesis}), we see an upward trend in the [X/Fe] beyond the knee -- which we have coined an elbow -- and for light and alpha-elements, we notice a decreasing trend afterwards (classic knee). For elements near the iron peak, we interpret this trend as the point at which SNe~Ia start to make a significant impact on chemical enrichment of the stellar population under study, considering that most of these elements are readily synthesised during SN~Ia explosions \citep{Thielemann2002}. Furthermore, we also consider the decreasing trend of alpha and light elements as the point where core-collapse supernovae has reached their peak, consequently leading to a decrease in their [X/Fe] as most of these elements are produced through explosions of massive stars \citep{Arnett1995, Iwamoto1999, Kobayashi2011}.
We conclude by noting again that the knee-age effect from this work is an indication of the variation in star formation rate due to the differences between older and younger stellar populations. This also suggests that there is a change in how gas is accreted onto the younger stellar sequence, which is influenced by the presence of the older sequence.

\subsection{How does our new selection criterion improve the selection of the thick and thin disc stars?} \label{sec:discussion_improvement_seleciton}
\begin{figure}
	\includegraphics[width=\columnwidth]{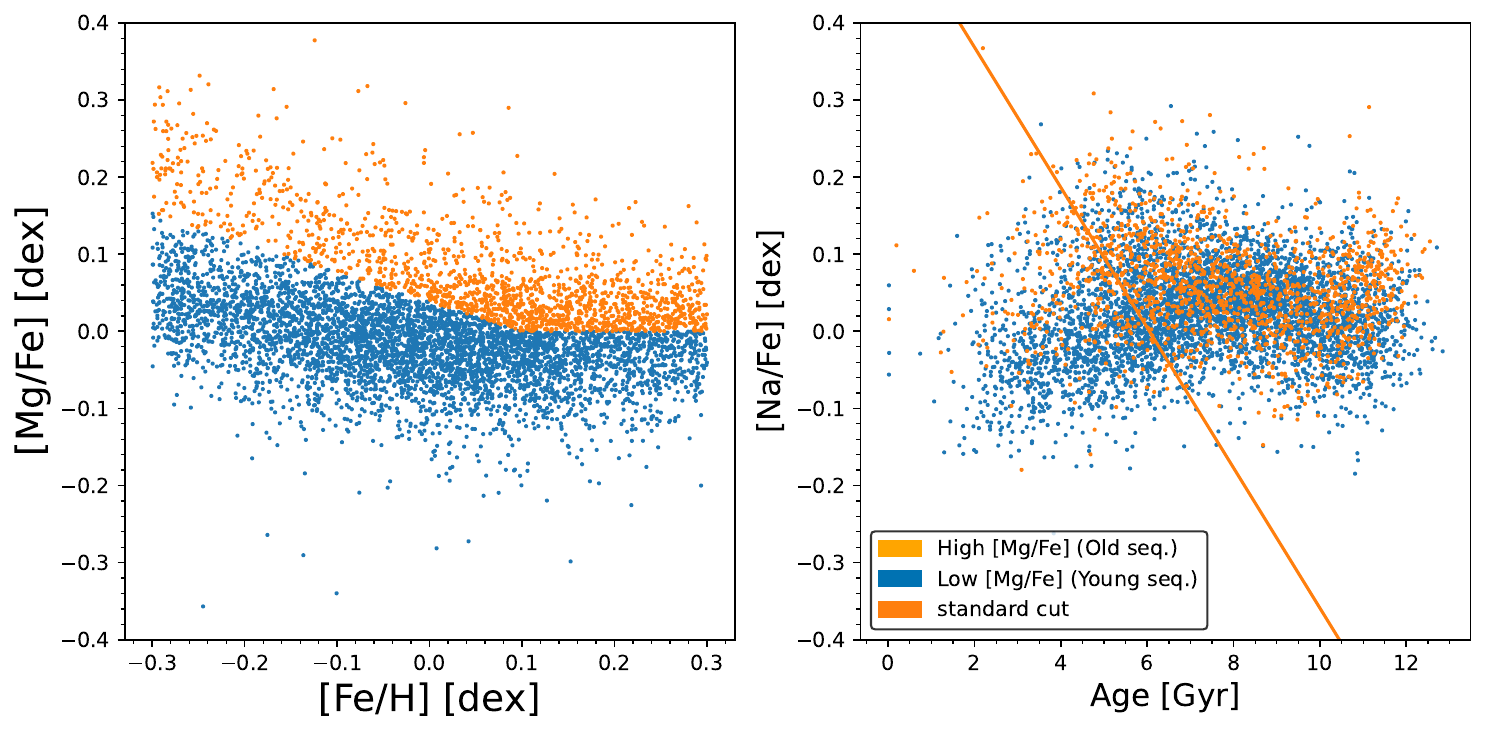}
    \caption{Visualisation of the selection difference of old and young disc stars with the method by \citet{Hayden2017}.
    Panel a) shows the plane of [Fe/H] vs. [Mg/Fe] that was used by \citet{Hayden2017} to select thin and thick disc stars.
    Panel b) shows that the same stars are less clearly separated in the [Na/Fe]--age plane} we propose in our study. Thick disc stars are orange, while a bluish colour represents the thin disc stars. 
    \label{fig:hayden}
\end{figure}

One of our motivations was to try and identify a less contaminated selection method of thin and thick disc stars in the high-metallicity regime, where stars significantly overlap in their dynamics and alpha enhancement. To this end, we have reproduced the linear selection in [Fe/H] vs. [Mg/Fe] used by \cite{Hayden2017} in the [Fe/H] vs [Mg/Fe] plane and we show where these stars are in the Age-[Na/Fe] plane in Fig. \ref{fig:hayden}. 

\begin{figure}
	\includegraphics[width=\columnwidth]{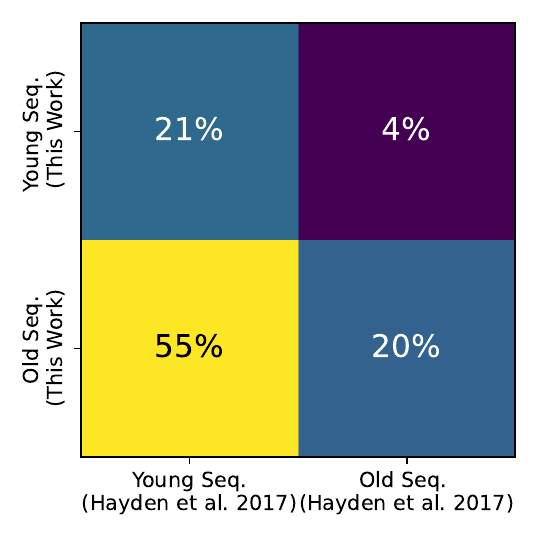}
    \caption{Confusion matrix of the selection of thin and thick disc by \citet{Hayden2017} via low and high [Mg/Fe] (x-axis) and our newly suggested selection method of young and old sequence disc stars via the age-[Na/Fe] plane (y-axis).} 
    \label{fig:confusion_matrix}
\end{figure}


\begin{figure*}
	\includegraphics[width=\textwidth]{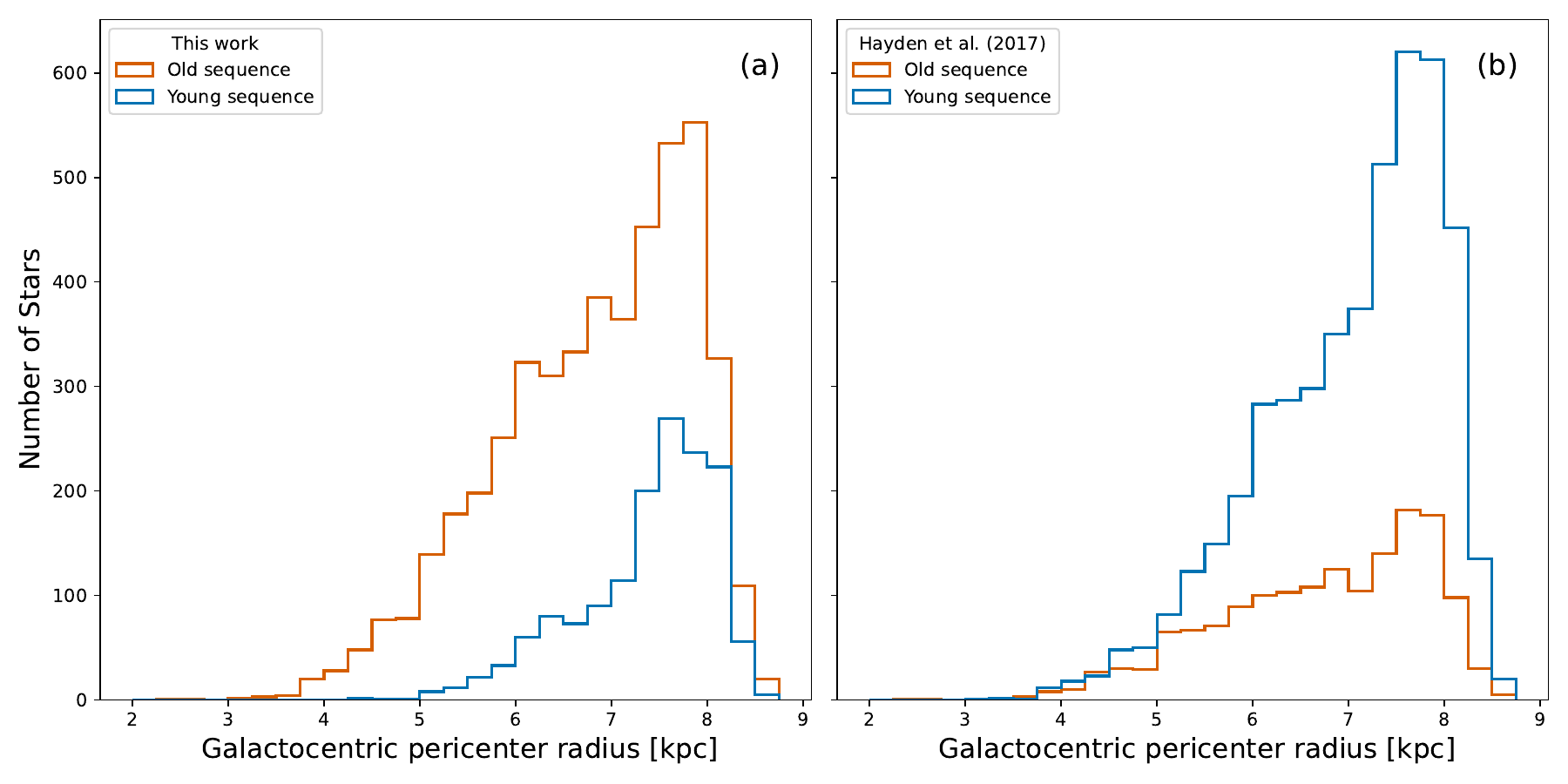}
    \caption{{Galactocentric pericenter radius distribution of the stars in our sample taken from the value-added catalogues presented in \citet{Buder2021} and based on \citet{GaiaCollaboration2021}. The Old and Young sequences have been separated by using the age-[Na/Fe] selection presented in this work in panel (a), and the selection presented by \citet{Hayden2017} in panel (b). A significant number of the old sequence stars in our work have smaller Galactocentric pericentre radii} in comparison to the young sequence stars, 
    which contrasts the \citet{Hayden2017} [Mg/Fe] based selection where both sequences of stars between 3--5 kpc display significant overlap.}
    \label{fig:pericenter_radius}
\end{figure*}
 
To compare how similar our new selection method is to the selection criteria of \citet{Hayden2017}, we construct a confusion matrix (see Fig.~\ref{fig:confusion_matrix}). We see that 59\% of the stars are attributed to opposite populations. Most notably, \citet{Hayden2017} would allocate 55\% (i.e., 3415 stars) of the old sequence as defined in our work as thin disc stars when using [Mg/Fe] criteria. Also, 4\% of young sequence stars (248 stars) as described in this work would be categorised as thick disc stars when using the [Mg/Fe] selection method. In all, only 41\% are allocated to the same populations (i.e., 21\%  thin and 20\% thick disc stars).

To further check as to whether our selection method results in less contaminated old and young sequences than the method of \citet{Hayden2017}, we also compare the orbital peri-galactocentric distances of the stars from both works. We indeed see a difference in the two populations: we find the older sequence stars are found to have smaller peri-galactocentric distances, e.g. distances closer to the Galactic centre, on average, compared to the younger sequence. 
We can clearly see a distinction between the old and young sequences, especially between 3--5 kpc from the histogram plot in Fig. \ref{fig:pericenter_radius} (panel a), where both sequences are distinctly differentiated in our Na-based selection compared to the Mg-based selection where they experience more overlap (Fig. \ref{fig:pericenter_radius}, panel b). We find a stronger correlation (supporting the notion that old sequence stars are more readily found closer to the inner Galaxy) than \citet{Hayden2017}. This significant difference in the allocation of stars to the two populations can have critical implications for the study of abundance trends in the Milky Way -- most notably the high-metallicity regime, which is more dominant in the inner Galaxy \citep{Matteucci1989, RecioBlanco2014}.

\section{Conclusions} \label{sec:conclusion}
In this study, we have presented an analysis using data from GALAH DR3 and quantified the abundances of elements at intermediate ages (8--5 Gyr) for solar-type stars (as defined by \citealt{Nissen2020}). We also presented abundance trends as a function of stellar age for light elements, alpha elements, odd-Z elements, iron peak elements, and neutron capture elements. We offered possible explanations for these elemental abundance trends where feasible.
There are two main outcomes of this work:
\begin{itemize}
\item Our selection function (Eq.~\ref{eq:line_age_na_fe}) allows us to identify age-abundance trends for the youngest thick disc (young end of the old sequence) stars that are challenging current nucleosynthetic ideas. This is most striking for Na, which shows an increase of $0.0260 \pm 0.0009\,\mathrm{dex/Gyr}$
 for old sequence stars between 8--5 Gyr (see Fig.~\ref{fig:combined_slopes_by_element}). We also present trends for some elemental abundances (relative to iron) as a function of age (see Tab. ~\ref{tab:age_percentiles}) that can be used for improving  Galactic chemical evolution models. 
\item We have shown that at the same age and metallicity, [Na/Fe] is a better tracer than [Mg/Fe] vs [Fe/H] to disentangle the old and young sequence disc stars. Our new approach using sodium allows us to better separate stars with similar [Fe/H] vs [Mg/Fe] into the old and young sequences, respectively (Fig.~\ref{fig:hayden}), with only 41\% of stars being allocated to the same populations in both studies (Fig.~\ref{fig:confusion_matrix}). 
\end{itemize}

\noindent In conclusion, we have shown that the unexpected behaviour of Na-enhancement at intermediate ages is a powerful discriminant to chemically separate the old and young sequence for solar-type stars in the Galactic disc. 
\section*{Acknowledgements}

We acknowledge the traditional owners of the land on which the AAT stands, the Gamilaraay people. We pay our respects to elders past and present and are proud to continue their tradition of surveying the night sky in the Southern Hemisphere. 
The authors acknowledge fruitful discussions with colleagues, including Alex Kemp, Benoît Côté, Simon Murphy, Giulia Cinquegrana, Philipp Eitner, and Nicholas Storm. 
This work was supported by the Australian Research Council Centre of Excellence for All Sky Astrophysics in 3 Dimensions (ASTRO 3D) through project number CE170100013. Part of this research was funded through AJR's Australian Research Council Future Fellowship FT170100243 and SB's Australian Research Council DECRA Fellowship DE240100150.

\section*{Facilities}

\textbf{AAT with 2df-HERMES at Siding Spring Observatory:}
The GALAH Survey is based on data acquired through the Australian Astronomical Observatory, under programs: A/2013B/13 (The GALAH pilot survey); A/2014A/25, A/2015A/19, A2017A/18 (The GALAH survey phase 1), A2018 A/18 (Open clusters with HERMES), A2019A/1 (Hierarchical star formation in Ori OB1), A2019A/15 (The GALAH survey phase 2), A/2015B/19, A/2016A/22, A/2016B/10, A/2017B/16, A/2018B/15 (The HERMES-TESS program), and A/2015A/3, A/2015B/1, A/2015B/19, A/2016A/22, A/2016B/12, A/2017A/14, (The HERMES K2-follow-up program). This paper includes data provided by AAO Data Central (datacentral.aao.gov.au).
\textbf{\textit{Gaia}: } This work has made use of data from the European Space Agency (ESA) mission \textit{Gaia} (\url{http://www.cosmos.esa.int/gaia}), processed by the \textit{Gaia} Data Processing and Analysis Consortium (DPAC, \url{http://www.cosmos.esa.int/web/gaia/dpac/consortium}). Funding for the DPAC has been provided by national institutions, particularly the institutions participating in the \textit{Gaia} Multilateral Agreement. 
\textbf{Other facilities:} This publication makes use of data products from the Two Micron All Sky Survey \citep{Skrutskie2006} and the CDS VizieR catalogue access tool \citep{Ochsenbein2000}.

\section*{Software}

The research for this publication was coded in \textsc{python} (version 3.7.4) and included its packages
\textsc{astropy} \citep[v. 3.2.2;][]{Robitaille2013,PriceWhelan2018},
\textsc{corner} \citep[v. 2.0.1;][]{corner},
\textsc{galpy} \citep[version 1.6.0;][]{Bovy2015},
\textsc{IPython} \citep[v. 7.8.0;][]{ipython},
\textsc{matplotlib} \citep[v. 3.1.3;][]{matplotlib},
\textsc{NumPy} \citep[v. 1.17.2;][]{numpy},
\textsc{scipy} \citep[version 1.3.1;][]{scipy},
We further made use of \textsc{topcat} \citep[version 4.7;][]{Taylor2005};

\section*{Data Availability}

The data used for this study is published by \citet{Buder2021} and can be accessed publicly via \url{https://docs.datacentral.org.au/galah/dr3/overview/}

\bibliographystyle{pasa-mnras}
\bibliography{references}
\label{lastpage}

\end{document}